\documentclass[final,5p,times,twocolumn,authoryear]{elsarticle}
\usepackage{amssymb}
\usepackage{amsmath}
\usepackage{amsthm}
\usepackage{xcolor}
\usepackage{csquotes}
\usepackage{float}
\usepackage{longtable}
\usepackage{multirow}
\usepackage[version=3]{mhchem} 
\usepackage{pdflscape}
\usepackage{lineno}
\usepackage{hyperref}
\usepackage{graphicx}
\defcitealias{shim18}{Shimonishi and Nakatani et al.}

\journal{Life Sciences in Space Research.}

\begin{document}

\begin{frontmatter}



\title{Binding energy distributions of alcohols, thiols, and their precursors on interstellar water ice surfaces}

\author[label1,label4]{Arghyadeb Roy} 
\ead{aroy.astro@gmail.com}
\author[label4]{Ankan Das} 
\ead{ankan.das@gmail.com}
\author[label10,label5,label4]{Milan Sil}  
\ead{milansil.astro@gmail.com}
\author[label2,label3,label4]{Prasanta Gorai}  
\author[label6,label7]{Kenji Furuya}  
\author[label8]{Naoki Nakatani} 
\author[label9]{Takashi Shimonishi} 

\affiliation[label1]
            {organization={Department of Chemical Sciences, Indian Institute of Science Education and Research Kolkata},
            city={Mohanpur},
            postcode={741246}, 
            state={West Bengal},
            country={India}}
\affiliation[label10]
            {organization={Institute of Astronomy, Department of Physics, National Tsing Hua University},
            city={Hsinchu},
            postcode={30013},
            country={Taiwan}}
\affiliation[label5]
            {organization={Univ. Grenoble Alpes, CNRS, IPAG}, 
            postcode={38000}, 
            city={Grenoble},
            country={France}}

\affiliation[label2]
            {organization={Rosseland Centre for Solar Physics, University of Oslo}, 
            postcode={PO Box 1029 Blindern}, 
            city={0315 Oslo},
            country={Norway}}

\affiliation[label3]
            {organization={Institute of Theoretical Astrophysics, University of Oslo}, 
            postcode={PO Box 1029 Blindern}, 
            city={0315 Oslo},
            country={Norway}}

\affiliation[label4]
            {organization={Institute of Astronomy Space and Earth Sciences},
            addressline={ P 177, CIT Road, Scheme 7m}, 
            city={Kolkata},
            postcode={700054}, 
            state={West Bengal},
            country={India}}

\affiliation[label6]
            {organization={National Astronomical Observatory of Japan},
            addressline={ Osawa 2-21-1}, 
            city={Mitaka},
            postcode={Tokyo 181-8588}, 
            country={Japan}}

\affiliation[label7]
            {organization={Department of Astronomy, Graduate School of Science, University of Tokyo},
            postcode={Tokyo 113-0033}, 
            country={Japan}}

\affiliation[label8]
            {organization={ Department of Chemistry, Graduate School of Science and Engineering, Tokyo Metropolitan University},
            addressline={1-1 Minami-Osawa}, 
            city={Hachioji},
            postcode={Tokyo 192-0397}, 
            country={Japan}}

\affiliation[label9]
            {organization={Institute of Science and Technology, Niigata University},
            addressline={Ikarashi-ninoncho 8050}, 
            city={Nishi-ku},
            postcode={Niigata 950-2181}, 
            country={Japan}}

\begin{abstract}
Binding energy (BE) is a critical parameter in astrochemical modeling, governing the retention of species on interstellar dust grains and their subsequent chemical evolution. However, conventional models often rely on single-valued BEs, overlooking the intrinsic distribution arising from diverse adsorption sites. In this study, we present BEs for monohydric alcohols, thiols, and their plausible precursors, including aldehydes and thioaldehydes. We incorporate a distribution of BEs to capture the realistic variation in adsorption strengths. The quantum chemical calculations provide a range of BE values rather than a single estimate, ensuring a more precise description of molecular diffusion and surface chemistry. The BE trend of analogous species provides qualitative insight into the dominant reaction pathways and key precursors that drive the formation of larger molecules under interstellar conditions. Oxygen-bearing species generally exhibit higher BEs than their sulfur analogues, primarily due to stronger interactions, further influencing molecular adsorption and reactivity. We implemented BE distributions in astrochemical models, revealing significant effects on predicted abundances and establishing a more accurate framework for future astrochemical modeling.
\end{abstract}



\begin{keyword}
Astrochemistry \sep Interstellar medium (ISM) \sep Molecular evolution \sep ISM: abundances \sep ISM: molecules  \sep ISM: molecular cloud
\end{keyword}

\end{frontmatter}

\section{Introduction} \label{sec1}
The study of astrochemistry has evolved significantly over the past several decades, shedding light on the molecular complexity of the interstellar medium (ISM) and the role of chemistry in various astrophysical environments. The detection of the first carbon-containing molecule, the methylidyne radical (CH), in 1937 \citep{Swings1937} marked the beginning of an era of molecular discoveries in space.
Since then, about 330 molecules, including neutrals, radicals, and ions, have been identified in the ISM and circumstellar shells\footnote{\url{https://cdms.astro.uni-koeln.de/classic/molecules}}. These discoveries have been made possible by advances in both ground-based and space-born observational facilities, enabling high-resolution spectral analysis of star and planet-forming regions.

A persistent challenge in astrochemistry is the missing sulfur problem. Observations of dense molecular clouds indicate that sulfur is significantly underrepresented in the gas phase by over two orders of magnitude relative to its expected solar abundance \citep{Ruffle1999,Wakelam2004, vant_Hoff2020}. This depletion suggests that sulfur atoms may be efficiently adsorbed onto dust grains, much like atomic oxygen, leading to the formation of hydrogenated sulfur species. However, recent studies suggest that the degree of sulfur depletion is not uniform across astrophysical environments, with translucent and dense clouds exhibiting depletion levels ranging from one to two orders of magnitude \citep{Fuente2023}. 

Although \ce{H2S} has been proposed as a major sulfur sink on dust grains \citep{Garrod2007,Escobar2011}, it has yet to be definitively detected in interstellar ices. So far, only a limited number of sulfur-bearing species, such as \ce{OCS} and tentatively \ce{SO2}, have been identified in the ice phase \citep{Yang2022,McClure2023,Rocha2024,Chen2024}, leaving the primary sulfur reservoirs largely unrevealed. Recent research suggests that organosulfur compounds (e.g., \ce{CS}, \ce{HCS}, \ce{H2CS}) and polysulfanes (e.g., $\rm{H_2S_n}$; $\text{n}=2,3$) may serve as key sulfur reservoirs, possibly becoming trapped in grain mantles or forming refractory materials that resist desorption into the gas phase \citep{Laas2019,Druard2012}. Additionally, in cosmic-ray-driven radiation chemistry, pure sulfur allotropes, primarily \ce{S8}, are the primary results of quick, non-diffusive reactions involving bulk radicals \citep{Shingledecker2020}.

In parallel, the formation of interstellar complex organic molecules (iCOMs) remains a critical area of interest in astrochemistry. 
{\color{black} Among them, monohydric alcohols, such as methanol (\ce{CH3OH}), have been widely detected in both the gas and solid phases of the ISM \citep[and references therein]{herb09,boog15} with a typical abundance of $10^{-7} - 10^{-6}$ respective to \ce{H2} in hot molecular cores \citep{Charnley1995}. Ethanol (\ce{C2H5OH}), the next member, is observed in hot cores with abundance ranging from $10^{-9} - 10^{-8}$ \citep[e.g.,][]{Sutton1995,MacDonald1996,Manigand2020}.}
More recently, the discovery of $n$-propanol (\ce{C3H7OH}) in the G$+0.693-0.027$ molecular cloud \citep{Jimenez-Serra2022} has expanded the known inventory of interstellar alcohols, with reported abundances of $(4.1 \pm 0.3) \times 10^{-10}$ for \textit{Ga}-\ce{C3H7OH} (\textit{Gt}-\ce{C3H7OH} by old nomenclature) and $(2.5 \pm 0.2) \times 10^{-10}$ for \textit{Aa}-\ce{C3H7OH} (\textit{Tt}-\ce{C3H7OH} by old nomenclature) conformers. 

Similarly, thiols, the sulfur analogues of alcohols, have been detected in various astrophysical environments. Methanethiol (\ce{CH3SH}) was first tentatively detected in Sgr B2 \citep{Turner1977} and later confirmed by \cite{Linke1979}, who found the \ce{CH3SH}/\ce{CH3OH} ratio to be consistent with the cosmic S/O ratio. \cite{Kolesnikova2014} suggested a tentative detection of Ethanethiol (\ce{C2H5SH}) in Orion KL. Recently, \cite{Rodriguez-Almeida2021} provided a definitive detection of the gauche isomer (\textit{g}-\ce{C2H5SH}) towards G$+0.693-0.027$ (same source as $n$-propanol) with an abundance of $\sim 3 \times 10^{-10}$, alongside \ce{CH3SH} with abundance $\sim 5 \times 10^{-9}$. Notably, they also claimed that the abundance ratios of SH-bearing species and their OH analogues exhibit similar trends with increasing molecular complexity.

Interestingly, while $n$-propanol has been confirmed in the ISM, its sulfur analogue, $n$-propanethiol (\ce{C3H7SH}), has yet to be observed, highlighting a significant gap in our understanding of interstellar sulfur chemistry. These discoveries provide essential clues about how sulfur is incorporated into organic molecules, offering potential solutions to the long-standing sulfur depletion problem.

\ce{CH3O} and \ce{CH2OH} are geometric isomers, generated from \ce{H2CO}, and are considered key precursors in forming monohydric alcohols (e.g., \ce{CH3OH}, \ce{C2H5OH}, and \ce{C3H7OH}). Similarly, \ce{CH3S} and \ce{CH2SH}, which originate from \ce{H2CS}, are proposed to act as precursors for monohydric thiols (e.g., \ce{CH3SH}, \ce{C2H5SH}, and \ce{C3H7SH}). Interstellar grains act as catalysts in this formation pathway. Determining the binding energy (BE) of these species is, therefore, crucial to understanding their role in such chemical pathways \citep{hase92,Das2008,Das2010,das16,das19,Das2011,sriv22,sil18,sil21,ghos22,Mondal2021}. In the absence of laboratory data, state-of-the-art quantum chemical approaches provide reliable estimates of the BE of interstellar species on grain surfaces. A crucial factor in modeling the abundance of these species in interstellar ice mantles is the accurate consideration of their BEs, as they determine the desorption temperature and residence time of molecules on grain surfaces, directly influencing their chemical reactivity and eventual gas-phase observability. However, many astrochemical models assume single BE values rather than a distribution, which can lead to significant deviations in predicted abundances \citep{Furuya2024}.

Since water (\ce{H2O}) is the dominant ice component in dense and cold regions of interstellar clouds, comprising $60-70\%$ of the ice along most lines of sight \citep{whit03,Gibb2004,boog15}, its interaction with other species plays a crucial role in determining their adsorption strengths. In dense and shielded regions, multiple layers of ice form through direct gas-phase adsorption or surface chemical processing. Although the precise structure of interstellar water ice remains uncertain, it is generally accepted that it forms as amorphous solid water (ASW) \citep{hama13}. Given this complexity, incorporating BE distributions rather than single values is essential for a more accurate representation of molecular desorption and ice chemistry in astrochemical models.

\begin{figure}
    \centering
    \includegraphics[width=1.0\linewidth]{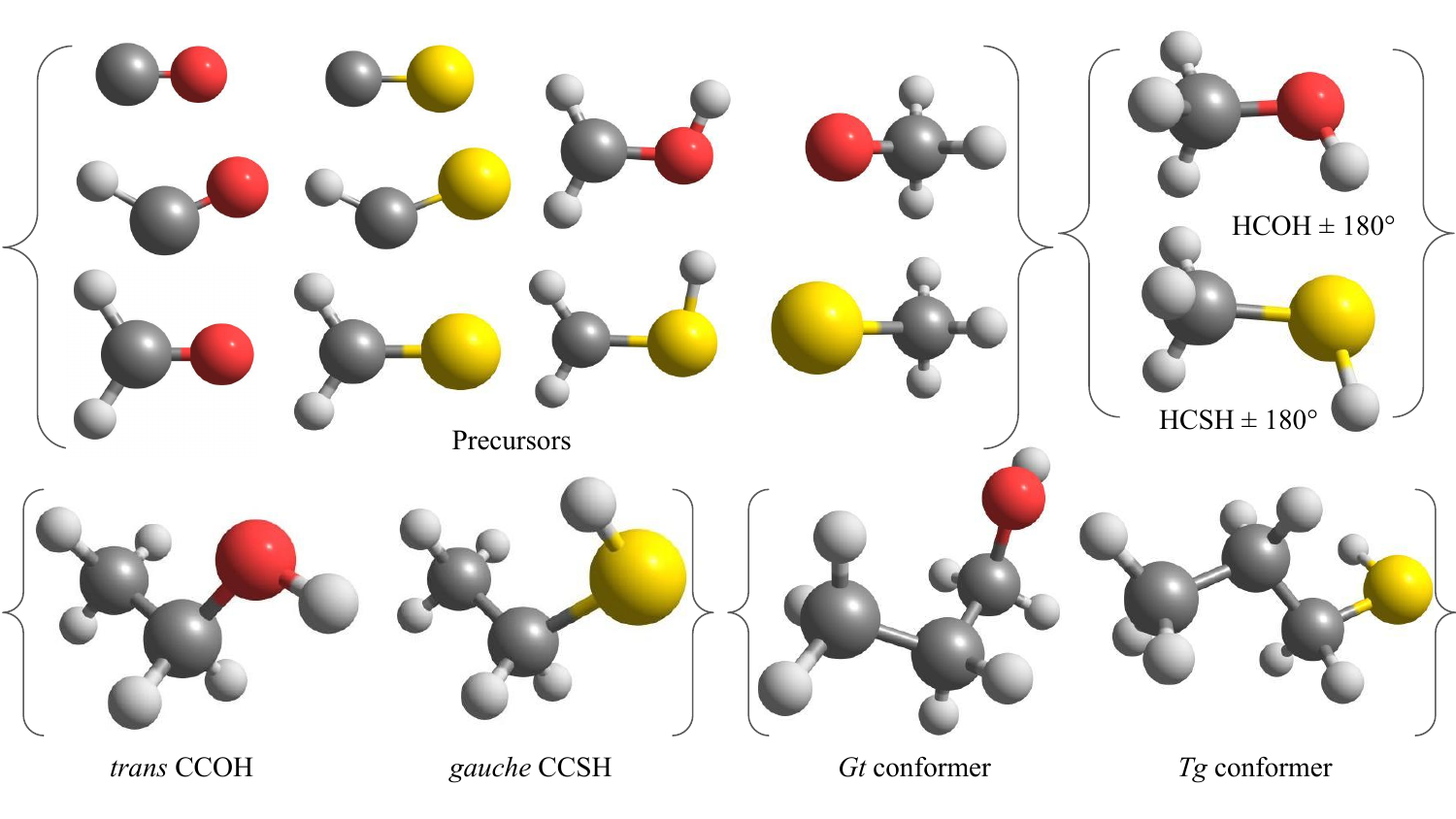}
    \includegraphics[width=0.6\linewidth]{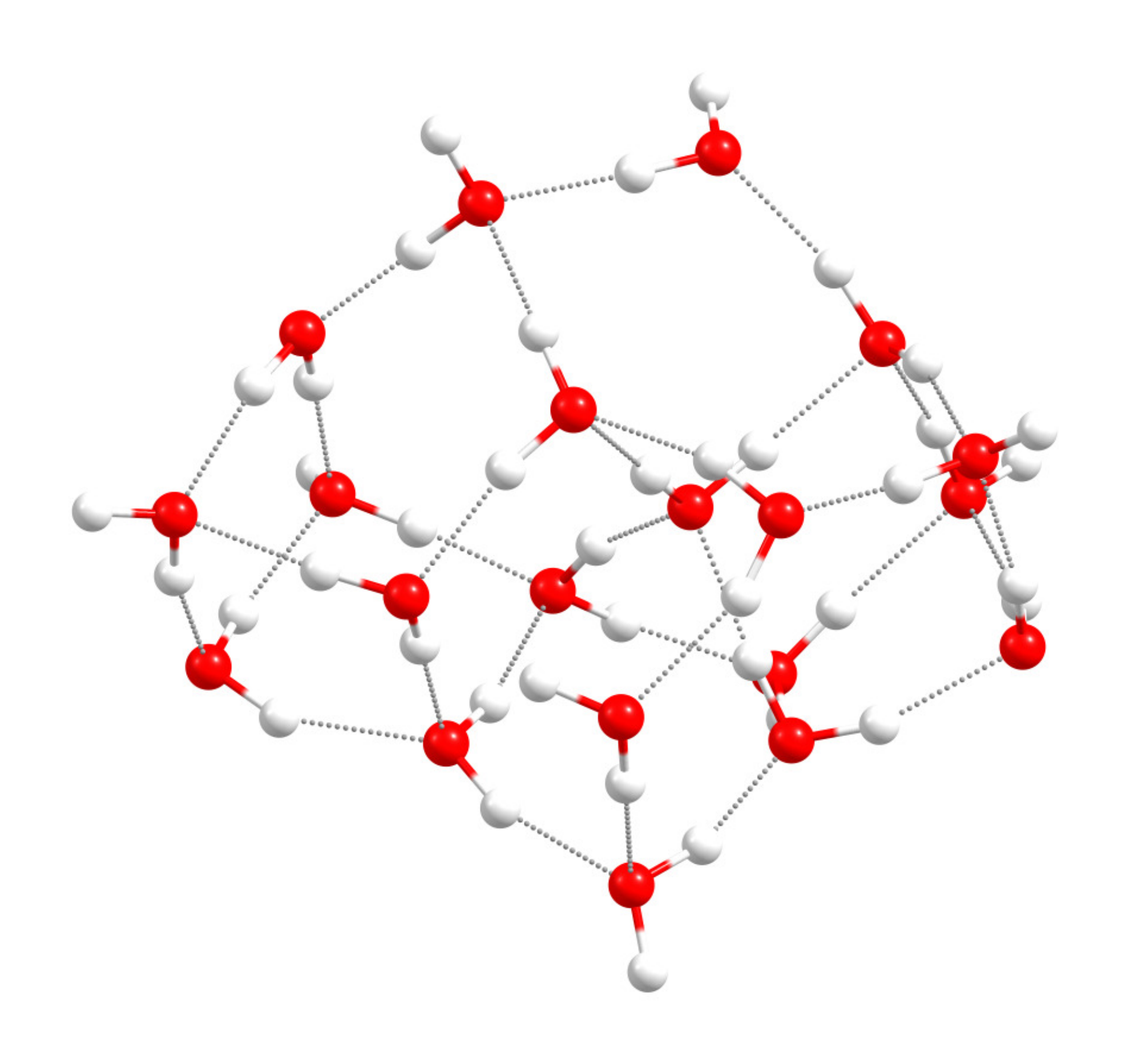}
    \caption{\color{black}The most stable structures of formaldehyde, thioformaldehyde, and their corresponding radicals as precursors, along with the monohydric alcohols and thiol analogues (\textit{top}) and the most stable ASW $\rm{[H_2O]_{20}}$ cluster structure (\textit{bottom}). The other ASW clusters with their relative energies are available in the appendix (available on \href{https://doi.org/10.5281/zenodo.13388691}{Zenodo}) of \cite{sil2024}. Red, white, grey, and yellow balls represent Oxygen, Hydrogen, Carbon, and Sulfur atoms, respectively.}
    \label{fig:species}
\end{figure}

{\color{black} In this context, numerous computational studies have explored the various ways molecules can adsorb on ASW surfaces, highlighting the remarkable diversity of these environments (\citealp{Sil2017,wake17,das18};~\citetalias{shim18},~\citeyear{shim18};~ \citealp{Ferrero_2020,Duflot2021,Sameera+2021,Perrero2022,EnriqueRomero_2022,Bovolenta2022,Piacentino2022,Hendrix2024,Martínez-Bachs_2024}).
 A new method for computing BE distributions has recently been introduced \citep{Germain2022,Tinacci2022,Tinacci2023,Bariosco2024,Bariosco2025}. These studies represent a significant improvement over previous approaches, as they conducted an in-depth analysis of the unbiased generation of water cluster models, as well as the unbiased sampling of binding sites for arbitrary molecules or radicals on generic cluster models of any composition, not limited to water.
Recently, \cite{sil2024} estimated BEs of seven diatomic radicals on \ce{[H2O]20} ASW clusters and evaluated the implications of the computed BEs for astrochemical models.}

This work presents BE range distributions for monohydric alcohols, thiols, and their plausible precursors, on ASW mimicking interstellar grain surfaces, addressing critical limitations in conventional astrochemical models that rely on single-valued BEs. We integrate multiple BE distributions into astrochemical modeling, building on the approach of \cite{Furuya2024} but applying it with a more realistic distribution. Additionally, we present BE values for new species that have not been previously studied on ASW, allowing us to investigate their impact on the abundance profiles of interstellar alcohol and thiol-related molecules in dense molecular cloud environments. The findings emphasize that incorporating BE distributions significantly impacts abundance predictions, underscoring the necessity of this approach for improving astrochemical simulations.

The remainder of this paper is organized as follows: Section \ref{sec2} describes the computational details adopted to estimate the BEs of the analogous species; Section \ref{sec3} includes our estimated BE results and their comparison with previous literature; Astrochemical modeling and its implications are discussed in Section \ref{sec4}; we conclude in Section \ref{sec5}.


\section{Computational Details and Methodology} \label{sec2}

The BE is often viewed as a local property arising from electrostatic interactions between the adsorbate and substrate. It is positive for a bound adsorbate and is defined as the energy difference between the separated adsorbate and substrate and the adsorbed complex.

In this work, the methodology proposed in \cite{sil2024} is considered to estimate the BE of both closed and open-shell species. Nine \ce{[H2O]20} ASW clusters (ASW 1–9), serve as adsorbents {\color{black}\citep[see appendix of][available on \href{https://doi.org/10.5281/zenodo.13388691}{Zenodo}]{sil2024}}. These were derived from crystalline ice made using the TIP3P model, heated to 300 K using MD simulations, and then rapidly cooled to 10~K {\color{black} (\citetalias{shim18},~\citeyear{shim18}). Based on the species and the available dangling-H ($d$-H) and dangling-O ($d$-O) positions on the ASW \ce{[H2O]20} cluster surface, the accessible binding sites are selected. Given its abundant binding sites, the \ce{[H2O]20} clusters seems like a good option for studying the BE.}

Quantum chemical computations were performed using \textsc{Gaussian 09} suite \citep{fris13}. Structures were optimized at the DFT $\omega$B97X-D/6-311+G(d,p) level, which includes long-range dispersion corrections and is suitable for radical systems. {\color{black} Harmonic vibrational frequency (unscaled) analysis confirmed local minima}, and zero-point energy (ZPE) corrections were applied. Basis Set Superposition Error (BSSE) was corrected using the CounterPoise method, where the corrected BE is obtained by subtracting BSSE from the uncorrected BE \citep{boys70}.

Geometry optimizations of the complicated structures are carried out after manually {\color{black} joining the species with the $d$-H/$d$-O using a weak H-like bond. The $d$-H atom of the \ce{[H2O]20} cluster is always preferred to be connected to the atom of the adsorbate with the greater electronegativity (in Pauling scale, H: 2.20; C: 2.55; S: 2.58; O: 3.44) by adhering to the electrostatic complementarity principle and for $d$-O, its the hydrogen atom connected to the most electronegative atom (here O and S)}. This method uses the increased electronegativity of an atom to draw electrons or electron density towards it, strengthening dipole interactions and producing more stable adsorption arrangements. {\color{black}All of the representative structures' coordinates are openly accessible in GitHub\footnote{\url{https://github.com/milansil/Binding-energy-distribution-data}} repository.}

\begin{figure}
    \centering
    \includegraphics[width=1.0\linewidth]{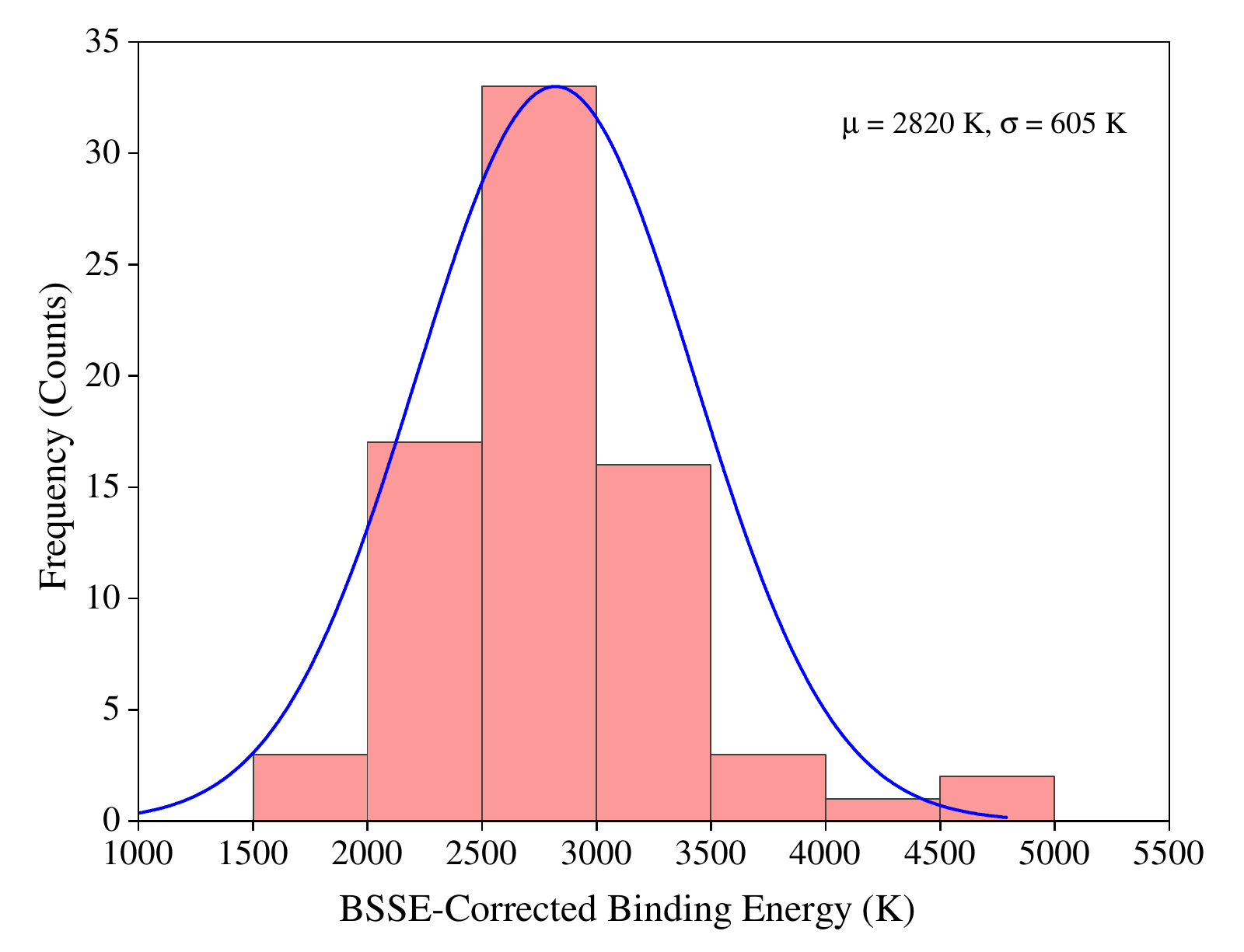}
    \caption{Gaussian distribution for the $+$BSSE BE values of \ce{H2CO} over the $75$ $d$-H binding sites across the nine ASW \ce{[H2O]20} clusters.}
    \label{fig:+BSSE_BE}
\end{figure}


\section{Results and Discussions} \label{sec3}
\subsection{Binding Energy}
\begin{table*}
\caption{BE calculations of species and their comparison.} \label{tab:BE}
\resizebox{\linewidth}{!}{\begin{tabular}{cccccccccc}
\hline
\hline
{Sl.} & {Species} & {Formula} & {Ground state} & \multicolumn{5}{c}{BE in Kelvin} \\
\cline{5-10}
{ No.} &&&{spin multiplicity} & {PENTEADO $^{(a)}$} & {DAS $^{(b)}$} &{UMIST $^{(\dagger)}$} &{KIDA $^{(\star)}$}& {Experimental} & {This work} \\
\hline
1. & Carbon monoxide & \ce{CO} & Singlet & $1100 \pm 250 $& $1263$ & $1150$ &$1300 \pm 390^{(c)}$; $1150^{(d)}$; $ 1390^{(k)}$  & $828 \pm 28^{(e)}$; $1180 \pm 20^{(f)}$; $1419^{(g)}$& {\color{black}$ 955 \pm 136 $}\\
2. & Formyl radical &\ce{HCO} & Doublet & $1355 \pm 500$& $1857$ & $1600$ & $2400 \pm 720^{(c)}$; $1600^{(d)}$ & -- & {\color{black}$ 2705 \pm 823 $}\\
3. & Formaldehyde & \ce{H2CO} & Singlet & $3260 \pm 60$ & $3242$ & $2050$ & $4500 \pm 1350^{(c)}$; $2050^{(d)}$ & $3259 \pm 60^{(h)}$ & {\color{black}$3696 \pm 663$} \\
4. & Methoxy radical & \ce{CH3O}  & Doublet & $2655 \pm 500$ &-- & $5080$ & $4400 \pm 1320^{(c)}$&-- & {\color{black}$ 3616 \pm 635 $}\\
5. & Hydroxymethyl radical & \ce{CH2OH} & Doublet& $2170 \pm 500$ & $4772$ & $5084$ & $4400 \pm 1320^{(c)}$; $5084^{(d)}$&-- & {\color{black}$5354 \pm 2111 $}\\
6. & Methanol & \ce{CH3OH}  & Singlet & $3820 \pm 135$ & $4368$  & $4930$& $5000 \pm 1500^{(c)}$; $5534^{(d)}$; $6621^{(k)}$  & $5410^{(i)}$& {\color{black}$5178 \pm 1343 $}\\
7. & Ethanol & \ce{C2H5OH}  & Singlet & $3470 \pm 500$  &  -- & $5200$ & $5400 \pm 1620^{(c)}$; $6584^{(d)}$ & $3127 - 7108^{(j)}$& {\color{black}$5383 \pm 1546$} \\
8. & $n$-propanol & \ce{C3H7OH}  & Singlet & --  & -- &  --& -- &--& {\color{black}$ 5626 \pm 1288 $}\\
\hline
\hline
9. & Carbon monosulfide & \ce{CS} & Singlet & $1800 \pm 500$ & $2217$ & $1900$& $3200 \pm 960^{(c)}$; $1900^{(d)}$; $4199^{(k)}$& -- &{\color{black}$ 2439 \pm 152 $} \\
10, & Thioformyl radical & \ce{HCS} & Doublet & $2000 \pm 500$ & $2713$ & $2350$ & $2900 \pm 870^{(c)}$; $2350^{d}$ & -- & {\color{black}$ 2074 \pm 593 $}\\
11. & Thioformaldehyde & \ce{H2CS} & Singlet & $2025 \pm 500$ & $3110$ & $2700$ & $4400 \pm 1320^{(c)}$; $2700^{(d)}$ & -- & {\color{black}$ 3307 \pm 740$}  \\
12. & Methanethiolate & \ce{CH3S}  & Doublet & --  & -- & -- & $4200 \pm 1260^{(c)}$ &--& {\color{black}$ 3059 \pm 711$} \\
13. & Sulfanylmethyl & \ce{CH2SH} & Doublet& -- & -- & -- &  $3700 \pm 1110^{(c)}$&--& {\color{black}$ 2961 \pm 771 $}  \\
14. & Methanethiol & \ce{CH3SH}  & Singlet & -- & -- & -- &$4000 \pm 1200^{(c)}$ & $4640 \pm 170^{(l)}$& {\color{black}$ 3652 \pm 974$} \\
15. & Ethanethiol & \ce{C2H5SH}  & Singlet & --  & -- & --& -- &-- &{\color{black}$ 3457 \pm 709 $}\\
16. & $n$-propanethiol & \ce{C3H7SH}  & Singlet & -- & -- & -- & --& --& {\color{black}$ 4445 \pm 1005$} \\
\hline
\end{tabular}}
{\small Ref$-$   $^{(a)}$\cite{Penteado_2017}; 
   $^{(b)}$\cite{das18}; 
      $^{(\dagger)}$UMIST: \url{http://www.udfa.net/};
   $^{(\star)}$KIDA: \url{https://kida.astrochem-tools.org/};
   $^{(c)}$\cite{wake17}; 
      $^{(d)}$OSU gas-phase database; 
$^{(e)}$ \cite{Noble2012a};
   $^{(f)}$ \cite{Collings2003};
   $^{(g)}$ \cite{Smith2016};
            $^{(h)}$\cite{Noble2012};
      $^{(i)}$\cite{Bahr2008};
       $^{(j)}$\cite{Jessica2024};
       $^{(k)}$\cite{Minissale2022};
              $^{(l)}$\cite{Narayanan2025}}
\end{table*}

We aim to calculate the BE of 16 species, as shown in the top panel of Fig.~\ref{fig:species}, to investigate the trend and impact of the distribution of BE on the formation of simple COMs through successive hydrogenation followed by their higher-order members.

First, we evaluated the BE of \ce{H2CO} across nine ASW clusters, summing up to 75 $d$-H binding positions. The BE values were then counterpoise-corrected to account for the BSSE. All the calculated values are presented in Table \ref{tab:all_H2CO}. To analyze the distribution of the BSSE corrected ($+$BSSE) BE values across the 75 $d$-H binding sites, we plotted their distribution using a bin width of $500$~K. As depicted in Fig.~\ref{fig:+BSSE_BE}, the data closely follows a Gaussian probability distribution function:
\begin{equation}
    \mathcal{G}(x,\mu,\sigma) = \frac{1}{\sqrt{2\pi\sigma^2}}\exp \bigg(-\frac{(x-\mu)^2}{2\sigma^2}\bigg)\,,
\end{equation}
where $\mu$ is the mean and $\sigma$ is the standard deviation. The fitted parameters for the distribution {\color{black} are $\mu=2820$~K and $\sigma=605$~K}.

Instead of averaging the BE over all 75 binding sites, we considered the average over the highest $d$-H binding sites in each of the nine clusters. This approach is more appropriate for estimating BE under low-temperature interstellar conditions, thus suggesting a Boltzmann distribution (i.e., a uniform distribution favouring the highest BE site below $10$~K), which holds for each local region. The resulting average $+$BSSE BE value for \ce{H2CO} {\color{black} is $3696$~K}.  

{\color{black} The BE values of the remaining 15 species were calculated and counterpoise corrected using binding sites of ASW(1) cluster (bottom panel of Fig. \ref{fig:species}), as it was found to be the most stable cluster ($\sim$ 2958 K more stable than the other eight clusters). In the case of the analogous alcohols and thiol species, we selected the most stable conformers of the molecules as the starting binding geometry, as previously determined by \cite{Gorai2017} (top panel of Fig. \ref{fig:species}). The highest BE value among the positions was selected for each of the species. For CO/CS, HCO/HCS, \ce{H2CS} and \ce{CH3O/CH3S} only the seven $d$-H sites are chosen as adsorption sites; whereas for \ce{CH2OH/CH2SH} and alcohol and thiols, since they can act as strong H-bond donor due to the functional group ($-$OH / SH), seven $d$-O along with the $d$-H sites are considered for BE evaluation.

To account for the impact of surface morphology variations, the highest BE value was then multiplied (except CO and CS) by a scaling factor: $1.172$ for $+$BSSE. This scaling factor was obtained by taking the ratio of the average BE to the highest BE for \ce{H2CO} on ASW(1), revealing a deviation of approximately $17\%$. The multiplication factor is derived to better understand the influence of different ASW surface morphologies on BE values. \ce{H2CO} is specifically chosen for this purpose as it is the first molecule from which the consecutive molecules and further alcohols form. Additionally, it has an experimental BE value for comparison. Another potential candidate could have been \ce{H2CS}, but it was not chosen due to the lack of an experimental value. 
The standard deviation in the BEs is obtained using the binding sites of the ASW(1) cluster and then multiplying them by the scaling factor of $2.385$ obtained through the similar methodology proposed in \cite{sil2024} to estimate the standard deviation for overall cluster morphologies. }

Notably, we evaluated the $+$BSSE BE of \ce{CO} and \ce{CS} at the seven $d$-H sites of the ASW(1) cluster and scaled them using a previously determined scaling factor of $1.177$ by \cite{sil2024}, as this factor was derived for diatomic radical species. Since \ce{CO} and \ce{CS} are diatomic, we applied this approximation to them. However, for larger and more branched species, a re-evaluation of the scaling factor was necessary, which we performed using \ce{H2CO}. The standard deviation for CO and CS was obtained using a scaling factor of $0.721$. {\color{black} The BE values for each sites and scaling is tabulated in Tables \ref{tab:CO_CS}$-$\ref{tab:dH_dO} and final BSSE-corrected scaled BE values and corresponding energy spread are presented in Table \ref{tab:BE}.}

\begin{figure}
    \centering
    \includegraphics[width=1.0\linewidth]{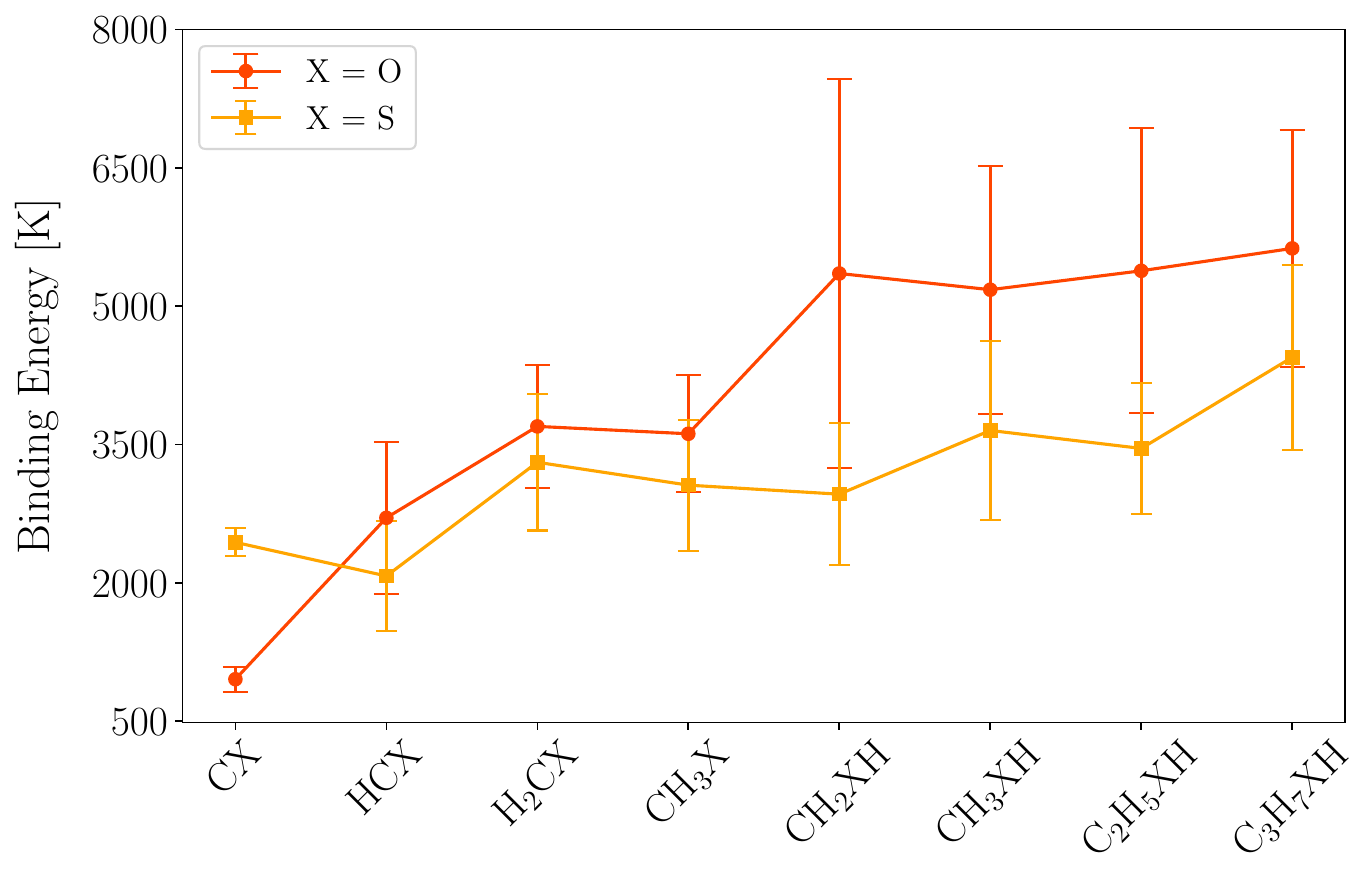}
    \caption{BE trend for the O$-$ and S$-$bearing analogue species on \ce{[H2O]20} clusters.}
    \label{fig:O&S}
\end{figure}

Fig.~\ref{fig:O&S} illustrates the BE trend for O$-$ and S$-$bearing analogous organosulfur compounds on \ce{[H2O]20} clusters. In general, there is an increasing BE trend for both cases on the ASW surface. CS exhibits a higher BE than CO, primarily due to its lower dispersive nature and significant dipole moment ($|\mu| \sim 1.96$ D). For other species, the BE of S$-$analogues are lower than that of their oxygen counterparts, likely due to stronger interactions with the water surface and the higher dipole moments of O$-$bearing species.

{\color{black} \ce{CH2OH} and \ce{CH3OH} display notably high BE values}, largely attributed to additional stabilization from other interactions with surfaces like H-bonding, suggesting that \ce{CH3OH} formation may proceed primarily via the \ce{CH3O + H} reaction, as their relatively low BE allows for surface diffusion, a conclusion also supported by microscopic kinetic Monte Carlo simulations from \cite{Simons2020}. In contrast, for \ce{CH3SH} formation, both \ce{CH3S} and \ce{CH2SH} can contribute similarly due to their comparable BE, indicating that multiple pathways may be viable under interstellar conditions.

\subsection{Comparison of BEs with literature values}

\begin{figure*}
    \centering
    \includegraphics[width=1.0\linewidth]{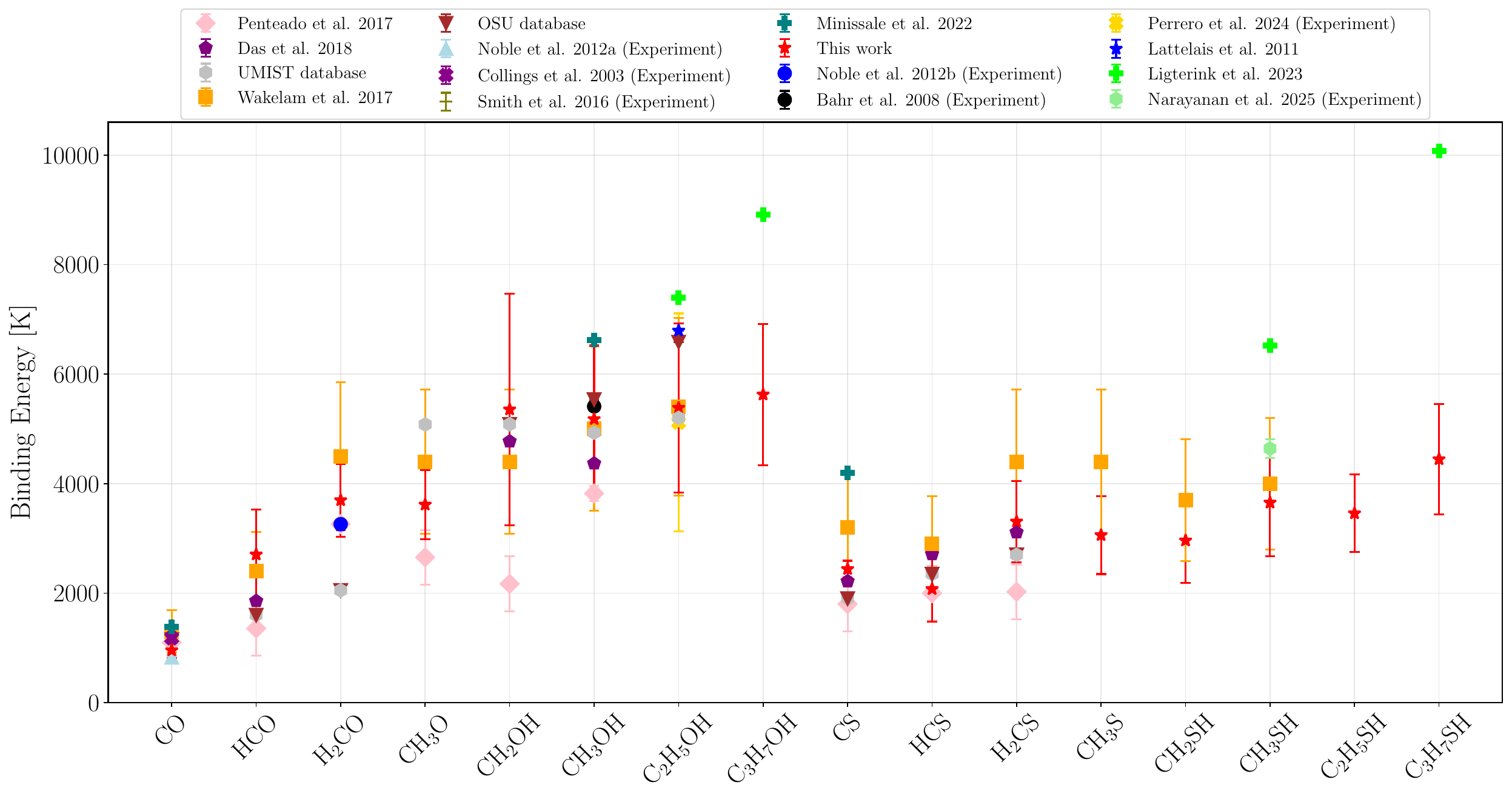}
    \caption{Comparison of reported BEs on water ice surfaces with our ZPE and BSSE corrected values.}
    \label{fig:comparison}
\end{figure*}

Experimentally measuring BEs is challenging, especially for radicals, due to their high reactivity during Temperature-Programmed Desorption (TPD), which complicates the evaluation of desorption parameters. The only experimentally available BE values are {\color{black}for \ce{CO}, \ce{H2CO}, \ce{CH3OH}, \ce{C2H5OH} and recently \ce{CH3SH}}. Consequently, benchmarking our computed values against experimental data is limited, and direct comparison with experimental results is not straightforward. Instead, we rely on existing theoretical reference values. Table \ref{tab:BE} presents our calculated BEs alongside available experimental, computational, and public database values. For clarity, these comparisons are visualized in Fig.~\ref{fig:comparison}, using a coherent dataset including recommendations from \cite{das18}, \cite{Penteado_2017}, other experimental and computational literature surveys, UMIST and the KInetic Database for Astrochemistry (KIDA).

\ce{CO}: {\color{black}  We report a BE value of  $955 \pm 136$~K for carbon monoxide, which is closer to experimental values obtained by \cite{Collings2003} ($1180 \pm 20$~K) and \cite{Noble2012a} ($828 \pm 28$~K) but lower than the value obtained by \cite{Smith2016} ($1419$~K). Notably, the value also overlaps with the BE range proposed by \cite{Penteado_2017} ($1100 \pm 250$~K), \cite{wake17} ($1300 \pm 390$), \cite{Bovolenta2022} ($1035 \pm 176$~K) and closer to UMIST and OSU database ($1150$~K). The value proposed by \cite{das18} ($1263$~K); \cite{Minissale2016} ($1300$~K) and \cite{Minissale2022} ($1390$~K) falls under the upper limit distribution. The value, however, is lower than the proposed value by \cite{Ferrero_2020} suggested a BE range of ($1109 -1869$~K for ASW and $1663$~K for crystalline). More recently, \cite{Groyne2025} evaluated the CO BE distribution using 266 binding sites; the resultant distribution is $1407 \pm 339$~K, with the median to be $1358$~K and \cite{Bulik2025} suggested an average BE value of $1560$~K.}

\ce{HCO}: {\color{black} We obtained a BE value of $2705 \pm 823$~K for formyl radical, which is in reasonable agreement with the estimates by \cite{wake17} ($2400 \pm 720$~K) and \citep{EnriqueRomero_2022} ($2466$~K on \ce{[H2O]18}) and the distribution closer to \ce{[H2O]33} value ($3536$~K). Our result is significantly higher than the values reported by \cite{Penteado_2017} ($1355 \pm 500$~K), \cite{Bovolenta2022} ($1317 \pm 378$~K), and the UMIST and OSU databases ($1600$~K). \cite{das18} reported a BE of $1857$~K and \citet{Duflot2021} reported a range of $1810 \pm 1149$~K, while \cite{Ferrero_2020} suggested a range of $1315-3081$~K for ASW and $2968$~K for crystalline ice, indicating surface dependence.}

\ce{H2CO}: We obtained a BE value of {\color{black}$3696 \pm 663$~K} for formaldehyde, which is in good agreement with the estimates by \cite{Penteado_2017} ($3260 \pm 60$~K), \cite{das18} ($3242$~K), and the experimental value reported by \cite{Noble2012} ($3259 \pm 60$~K). However, our result is significantly higher than the UMIST and OSU database value of $2050$~K. \cite{wake17} suggested an extrapolated value of $4500 \pm 1350$~K. \cite{Ferrero_2020} reported a BE of $5187$~K for crystalline ice and a range of $3071 - 6194$~K for ASW, which overlaps with our findings on the ASW surface. More recently, \cite{Bovolenta2022} reported a mean BE of $2970$~K and a maximum BE of $3800$~K.

\ce{CH3O}: We report an average BE for the methoxy radical of  {\color{black}$3616 \pm 635$~K}, which falls within the range of $4400 \pm 1320$~K proposed by \cite{wake17} using water monomer. This value is also consistent with the average BE of $3598$~K reported by \cite{Sameera+2021} for [\ce{H2O}]$_{162}$ ASW cluster but is higher than the average BE of $3249$~K on an icosahedral ($I_h$) water cluster. However, our calculated value is significantly higher than the recommended value of $2655 \pm 500$~K reported by \cite{Penteado_2017}, conversely lower than the proposed value in UMIST database ($5080$~K). Recently, \cite{EnriqueRomero_2022} calculated BEs for \ce{CH3O} as $3139$~K and $4582$~K on the \ce{[H2O]18} and \ce{[H2O]33} ASW ice models, respectively, assuming a single binding site. While insightful, this assumption oversimplifies the complexity of interstellar surfaces, where multiple binding sites exist, resulting in a distribution of binding energies. \cite{Bovolenta2022} reported a mean BE of $2274$~K and the maximum BE of $3343$~K. Notably, the BE obtained by our \ce{[H2O]20} ASW ice model falls within the values obtained by \cite{EnriqueRomero_2022}.

\ce{CH2OH}: We report a BE of {\color{black}$5354 \pm 2111$~K} for the hydroxymethyl radical, which is in good agreement with the UMIST and OSU database value of $5084$~K and falls at the higher end of the $4400 \pm 1320$~K range proposed by \cite{wake17}. Notably, \cite{wake17} suggested that the BEs of \ce{CH3O} and \ce{CH2OH} lie within the same range; however, our results demonstrate that this is not the case. Our calculated value is also significantly higher than the recommended value of $2170 \pm 500$~K proposed by \cite{Penteado_2017} and exceeds the $4772$~K reported by \cite{das18}. Similar to \ce{CH3O}, the BE of \ce{CH2OH} has also been evaluated by \cite{EnriqueRomero_2022}, who reported values of $5521$~K and $6170$~K on the \ce{[H2O]18} and \ce{[H2O]33} ASW ice models, respectively. A mean BE of $4451$~K for the major binding mode and $2670$~K for the minor binding mode and the maximum BE of $6594$~K was reported by \cite{Bovolenta2022}. \cite{Sameera+2023} recently reported a BE range of $3365-8007$~K, with an average value of $5686$~K considering eight binding position each from the $I_h$ and ASW cluster.

\ce{CH3OH}: Methanol is the first candidate in the list of monohydric alcohols. We obtained a BE value of {\color{black}$5178 \pm 1343$~K, which is in correspondence with the experimental value of $5410$~K reported by \cite{Bahr2008}; UMIST database value of $4930$~K and the OSU database value of $5534$~K and $5000 \pm 1500$~K proposed by \cite{wake17}. However, it is far higher than the proposed value of $3820 \pm 135$~K by \cite{Penteado_2017}, and our result is also notably higher than the computed value of $4368$~K reported by \cite{das18}. \cite{Ferrero_2020} reported a BE of $7385$~K on crystalline ice and a range of $3770-8618$~K on ASW surfaces, reflecting the variability of BEs across different surface types. Recently, \cite{Minissale2022} suggested a BE value of $6621$~K. Similar to \ce{CH2OH}, \cite{Bovolenta2022} found a mean BE of the major binding mode to be $3235$~K, whereas for the minor binding mode, the value is $2344$~K, with the highest to be $5331$~K. \cite{Sameera+2023} reported a BE range of $1741-8356$, with an average value of $4758$~K. Very recently, \cite{Bariosco2025} reported a BE range of $4255 \pm 1558$~K and \cite{Bulik2025} suggested an average BE value of $5040$~K. Both are in agreement with our findings.}

\ce{C2H5OH}: The second candidate in the list of monohydric alcohols is ethanol, which is considered the parent molecule of formic acid (\ce{HCOOH}) and other iCOMs, such as glycolaldehyde (\ce{HCOCH2OH}), acetic acid (\ce{CH3COOH}), and acetaldehyde (\ce{CH3CHO}) \citep{Skouteris2018}. Experimentally determining the BE of ethanol is challenging due to its ability to form multiple hydrogen bonds with water surfaces. This results in complex TPD profiles, which are often hindered by co-desorption phenomena and occasionally exhibit multiple desorption peaks \citep{Lattelais2011,Jessica2024}. For this reason, \cite{Burke2015} classified ethanol as a complex, water-like molecule that wets the ice surface, co-desorbs with water, and even influences the water crystallization process. \cite{Jessica2024} provided an experimental BE range of $3127-7108$~K for single adsorption on ASW surfaces and a theoretical estimation of $4450-5773$~K on ASW cluster. We obtained a BE value of {\color{black}$5383 \pm 1546$~K}, which overlaps with the UMIST database value of $5200$~K and is closer to the OSU database value of $6584$~K and the theoretical value of $6794$~K proposed by \cite{Lattelais2011}. Our result also falls within the range of $5400 \pm 1620$~K proposed by \cite{wake17}. Recently, \cite{Ligterink2023} suggested an overestimated BE value of $7400$~K through Redhead-Transition State Theory (TST) approximation.

\ce{C3H7OH}: To the best of our knowledge, we are the first one to rigorously derive the BE of $n$-propanol using a water cluster. We report a BE value of {\color{black}$5626 \pm 1288$~K}. Notably, \cite{Ligterink2023} suggested a BE value of $8909$~K with its own uncertainties, as stated previously in the case of ethanol. 

\ce{CS}: For carbon monosulfide, we report a BE value of {\color{black}$2439 \pm 152$~K}, which is closer to the values proposed by \cite{Penteado_2017} ($1800 \pm 500$~K) and \cite{das18} ($2217$~K) but lower than \cite{wake17} ($3200 \pm 900$~K) whereas the suggested value by UMIST and OSU database is $1900$~K. Recently, \cite{Jessica2022} reported a BE value of $2453$~K for crystalline and with a deviation of $909$~K for ASW surface. \cite{Minissale2022} suggested a high BE value of $4199$~K.

\ce{HCS}: We obtained a BE value of {\color{black}$2074 \pm 593$~K} for HCS, which is in good agreement with the estimates by \cite{Penteado_2017} ($2000 \pm 500$~K) and \cite{Jessica2022} ($2009 \pm 469$~K for ASW). However, our result is lower than the values reported by \cite{das18} ($2713$~K), \cite{wake17} ($2900 \pm 870$~K), and the UMIST and OSU databases ($2350$~K). \cite{Jessica2022} also reported a higher BE of $3318$~K for crystalline ice surface.

\ce{H2CS}: For thioformaldehyde, we obtained a BE value of {\color{black}$3307 \pm 740$~K}, which aligns well with the estimated $3110$~K reported by \cite{das18}. However, our result is higher than the values reported by \cite{Penteado_2017}, the UMIST and OSU database, which estimate the BE at $2025 \pm 500$~K and $2700$~K, respectively. \cite{wake17} extrapolated a water monomer BE value of $4400 \pm 1320$~K. More recently, \cite{Jessica2022} suggested a BE of $3692 \pm 859$~K for ASW and $4267$~K, suggesting good agreement with our findings.

\ce{CH3S}: We report an average BE for methanethiolate of {\color{black}$3059 \pm 711$~K}, which falls at the lower end of the range of $4400 \pm 1320$~K proposed by \cite{wake17}. Notably, \cite{wake17} estimated that \ce{CH3O} and \ce{CH3S} would have a similar range of BE distributions, an assessment that aligns well with our results.

\ce{CH2SH}: Similarly, for sulfanylmethyl, we obtained a BE value of {\color{black}$2961 \pm 771$~K}, which also falls in the BE range of $3700 \pm 1110$~K estimated by \cite{wake17}.

\ce{CH3SH}: We report a BE of {\color{black}$3652 \pm 974$~K} for methanethiol (methanol analogue), which lies within the range of $4000 \pm 1200$~K proposed by \cite{wake17}. \cite{Jessica2022} recently reported BE values of $4603$~K on crystalline ice and $3603 \pm 769$~K on ASW surfaces. \cite{Ligterink2023} proposed an overestimated BE value of $6522$~K. Very recently, \cite{Narayanan2025} reported an experimentally obtained value of $4640 \pm 170$~K.

\ce{C2H5SH}: For ethanethiol, too, we are the first to rigorously determine its BE, obtaining a value of {\color{black}$3457 \pm 709$} on \ce{[H2O]20} ASW surface.

\ce{C3H7SH}: \cite{Ligterink2023} suggested a BE value of $10070$~K for $n$-propanethiol; which are estimated based on the BE values obtained on Au(111) surface. We report a BE value of {\color{black}$4445 \pm 1005$~K} on \ce{[H2O]20} ASW surface.


\section{Astrochemical Modeling and Implications} \label{sec4}

\begin{figure}
    \centering
    \includegraphics[width=1.0\linewidth]{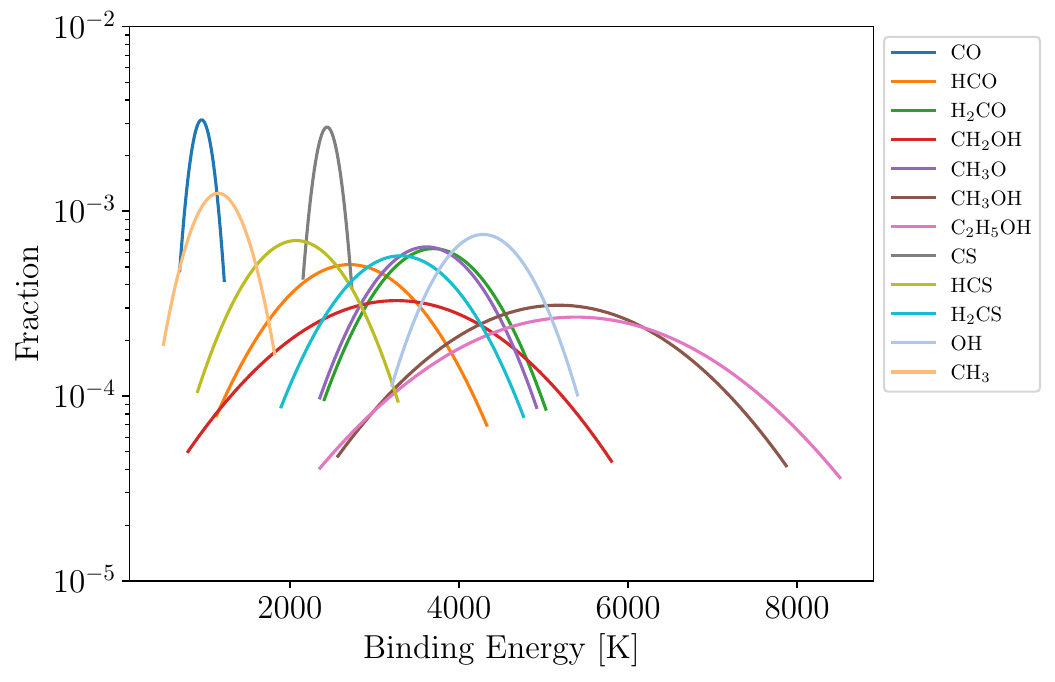}
    \caption{ {\color{black} The fraction of binding sites with binding energies. The binding energy distribution for each species is assumed to follow a Gaussian distribution based on the four parameters noted in Table \ref{tab:BE}: the mean binding energy, standard deviation ($\sigma$), and maximum and minimum binding energies considering 2$\sigma$ variations. }}
    \label{fig:BEdist}
\end{figure}

We utilize the {\scshape{Rokko}} chemical code \citep{furu15} to investigate the impact of our calculated BEs on the chemical model. While traditional rate equation methods typically concentrate on a single binding site, both laboratory and computational studies have demonstrated the presence of multiple binding sites. \cite{Furuya2024} developed a framework incorporating the BE distribution of various species within the rate equation approach. This method is computationally efficient and can be effectively applied to extensive gas-grain chemical networks. In this study, we analyze our computed BE distribution within the model, which is represented by a Gaussian distribution, as shown in Fig.~\ref{fig:BEdist}.

{\color{black} Previous astrochemical models have highlighted the complex nature of sulfur-bearing species due to their rich and varied chemistry. The reaction networks often involve uncertain gas-phase and grain-surface pathways, which makes accurate modeling challenging. Additionally, sulfur depletion and its redistribution among various reservoirs further complicate predictions of abundance. These complexities frequently result in significant discrepancies between models and observations \citep{char97,Wakelam2004, vida17, Gorai2017, Laas2019}.}
Since the reaction network for the formation of thiols (${\rm CH_3SH, \ C_2H_5SH, \ C_3H_7SH}$) and higher-order alcohols (${\rm C_3H_7OH}$) was not included in {\color{black} our present} chemical network, we have omitted them from the model.

\begin{figure}
    \centering
    \includegraphics[width=1.0\linewidth]{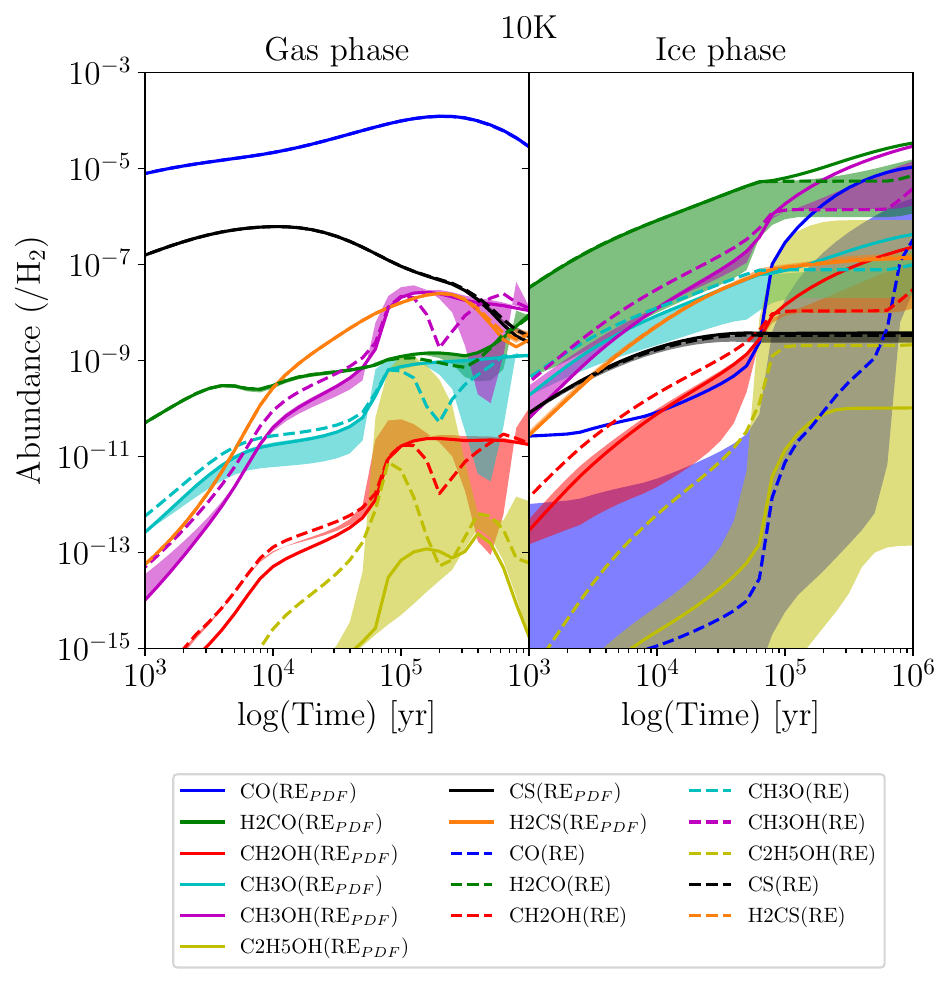}
     \includegraphics[width=1.0\linewidth]{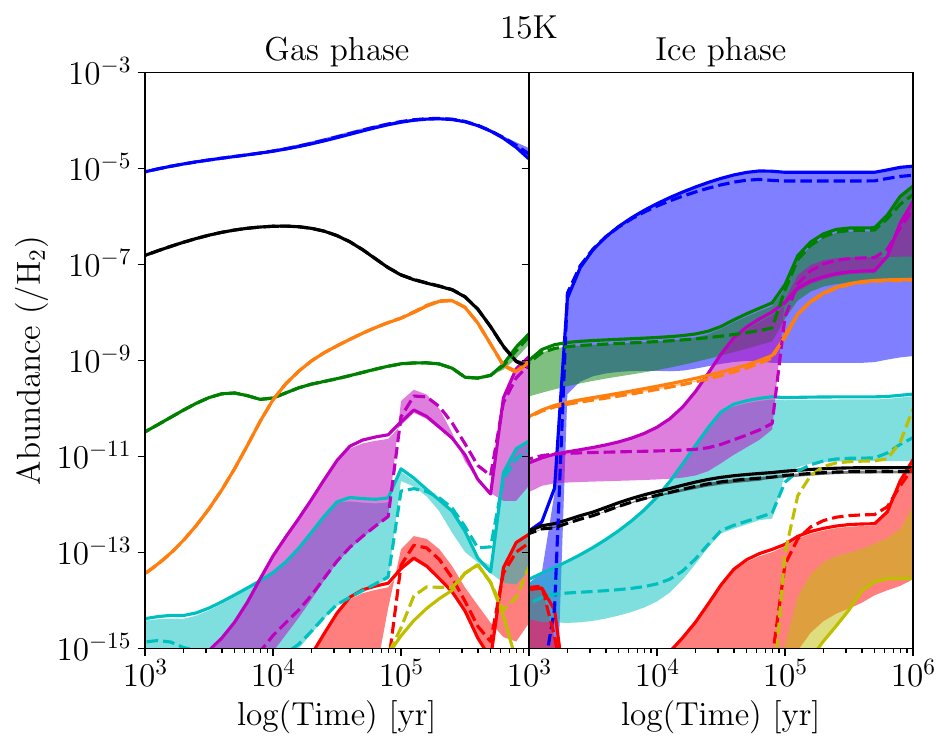}
    \includegraphics[width=1.0\linewidth]{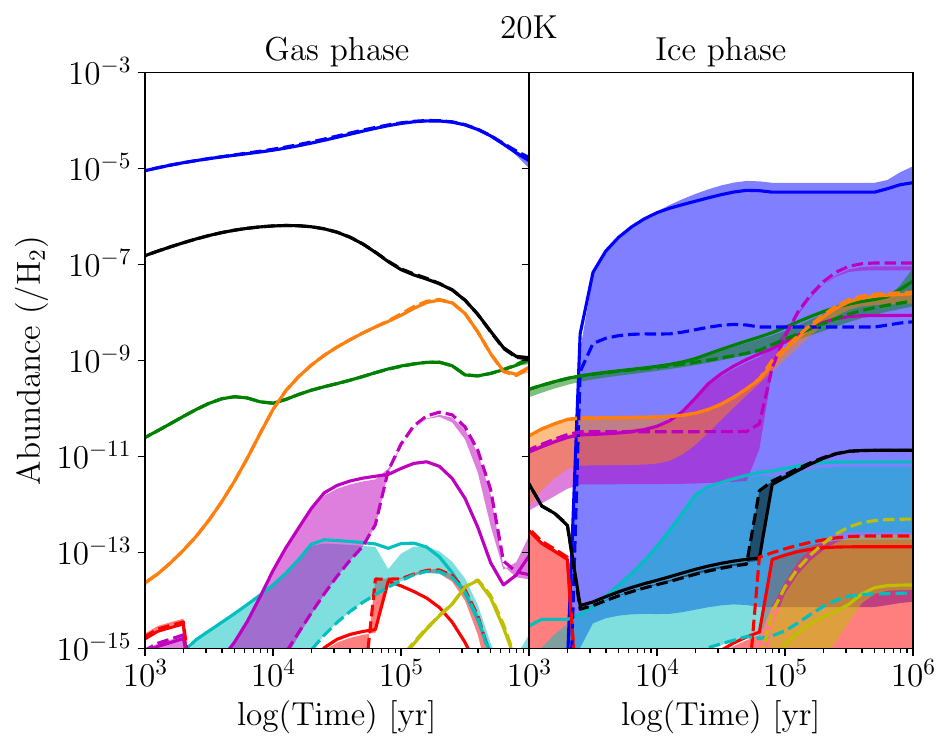}
      \caption{Time evolution of abundances obtained with the rate equation model (RE) and BE distribution model (RE$_{PDF}$) at 10~K (\textit{top}); 15~K (\textit{middle}) and 20~K (\textit{bottom}). The left panels of these figures show the evolution in the gas phase, whereas the right panels illustrate it in the ice phase, including both surface and bulk.}
    \label{fig:Chem}
\end{figure}

We consider a dark cloud model with a hydrogen density of \( n_\text{H} = 2 \times 10^4 \, \text{cm}^{-3} \), a visual extinction \( A_V = 10 \), a cosmic ray ionization rate, \( \zeta_{CR} = 1.3 \times 10^{-17} \, \text{s}^{-1} \) and initial abundances from \cite{mill22} to compare the differences between the traditional rate equation (RE) model and the rate equation model that incorporates a BE distribution framework, referred to as \( \text{RE}_{PDF} \) \citep{Furuya2024}. It is important to note that our \( \text{RE}_{PDF} \) model does not consider the BE distribution of all species; it only considers the BE distribution depicted in Fig. \ref{fig:BEdist}. We considered that gas and ice are well coupled and maintain a similar temperature.
A temperature of 10 K, 15 K, and 20 K (Fig. \ref{fig:Chem}) is considered to study the differences between the RE and RE$_{PDF}$ models.
The traditional RE model considers only single binding sites, using the mean BE values from the last column of Table \ref{tab:BE}. In contrast, the RE model also includes the minimum and maximum BE values obtained from our calculations. The shaded region in Fig. \ref{fig:Chem} represents the abundances derived from these minimum and maximum BEs.
We observed significant differences in the predicted abundances of ice-phase species between the RE and \( \text{RE}_{PDF} \) models, especially when compared to gas-phase species. However, the current \( \text{RE}_{PDF} \) model is not suitable for temperatures above 100 K, where most ice mantles begin to sublimate, thereby affecting gas-phase abundances more prominently. 

The RE model predicts a marginally higher estimation of methanol production compared to the RE$_{PDF}$ model. This finding contrasts with the results reported by \cite{Furuya2024}, which indicated that the RE model tends to underproduce methanol. Notably, the study considered the BE distribution of hydrogen atoms, a factor not addressed here. Given that hydrogenation is crucial for regulating surface coverage at lower temperatures, omitting this variation may be the reason for these differences. Table \ref{tab:BE} indicates that the mean BE for all the species we studied, except for CO, is relatively high, exceeding 2000 K.  Consequently, the abundance of methanol at low temperatures appears to be significantly influenced by the BE of CO.
{\color{black} Since the BE of CS and HCS are high (exceeds 2000 K), we did not observe any notable differences in the abundances of CS and \ce{H2CS}, as shown in Fig. \ref{fig:Chem}.}

\section{Conclusions} \label{sec5}

In this study, we calculated the BEs of 16 interstellar species on ASW surfaces, incorporating BSSE corrections and scaling factors to account for surface heterogeneity. Our results indicate that O$-$ and S$-$bearing analogues follow a general trend of increasing BE, with sulfur-containing species exhibiting lower values due to weaker interactions with the ASW surface. These findings emphasize the importance of accurately characterizing BE distributions, as they significantly impact astrochemical modeling. By incorporating BE distributions into rate equation (\(\text{RE}_{PDF}\)) models, we observed notable differences in ice-phase abundances compared to traditional fixed-BE rate equation (RE) models. The impact is particularly evident for volatile species such as CO and methanol, where surface binding strength directly influences desorption and subsequent chemical pathways. Our results highlight the necessity of including BE distributions to better constrain ice-phase abundances in astrochemical simulations.

We conducted a comparative analysis between the traditional rate equation model and a modified rate equation that includes the BE distribution of specific species. Our findings indicate that incorporating the BE distribution would have a significant impact on the abundance of gas-phase species beyond 100 K, where most of the ice constituents sublime.
 Currently, our RE$_{PDF}$ model is applicable within the low-temperature region. We are optimistic that our future RE$_{PDF}$ model, designed for the hot core region, will effectively address and enhance our understanding.
 We expect that deviations may become more pronounced at higher temperatures, where radical-radical reactions are likely to dominate surface coverage, and at elevated temperatures when most ice mantles start to sublimate.

Future studies will focus on refining the pre-exponential factor in desorption calculations, as well as extending the BE distribution approach to sulfur chemistry and other key astrochemical species for the high-temperature regime. Additionally, exploring the impact of surface morphology on BE variations will further improve the accuracy of astrochemical models and enhance our understanding of molecule-surface interactions in interstellar environments.

\section*{Acknowledgements}
The authors thank the anonymous referees for their valuable insights and constructive comments, which significantly improved the manuscript. This work was carried out while A.R. was a Master’s student at IISER Kolkata. A.R. gratefully acknowledges support from the DST-INSPIRE-SHE fellowship during this period of time. M.S. acknowledges the support provided by the National Science and Technology Council, Taiwan (Grant Nos. NSTC-111-2112-M-007-014-MY3, NSTC-113-2639-M-A49-002-ASP, NSTC-113-2112-M-007-027, and NSTC-114-2811-M-007-017).

\bibliographystyle{elsarticle-harv} 
\bibliography{references}


\appendix
\setcounter{table}{0}

\onecolumn

\begin{landscape}

\section{Calculated BE values for the species using $\rm{[H_2O]_{20}}$ ASW clusters and DFT-$\omega$B97X-D/6-311+G(d,p) level theory.}

  \begin{table}[H]
   \centering
   \caption{{\color{black}Calculated BE values for {H$_2$CO} considering 9 different ASW $\rm{[H_2O]_{20}}$ clusters. All values reported are in the Kelvin (K) unit.}}
  \phantomsection
   \label{tab:all_H2CO}
    \resizebox{\linewidth}{!}{\begin{tabular}{c c c c c c c c c c c c c c}
    \hline
    \hline
{Cluster} &{ BSSE} &\multicolumn{10}{c}{$d$-H binding sites} & \multicolumn{2}{c}{ Highest BE value} \\
\cline{3-12} 
{ No.} &{ $(-/+)$} &{$d$H1 } & { $d$H2 } & { $d$H3 } & { $d$H4 } & { $d$H5 } & { $d$H6 } & {$d$H7 } & { $d$H8 } &  { $d$H9 }&  { $d$H10 } & { $-$BSSE} & { $+$BSSE} \\
\hline
\multirow{2}{*}{ASW(1)}&$-$BSSE & 2976& 3040 & 3193& 2635& 3354& 2779&2541 &-- &-- &-- & \multirow{2}{*}{3354} &  \multirow{2}{*}{3154} \\
& $+$BSSE& 2772 & 2855 & 2816 & 2418 & 3154 & 2578 & 2352 &-- &-- &--  \\
\hline
\multirow{2}{*}{ASW(2)}&$-$BSSE  &2975 &2794 &2549 &4207 &2942 &2990 & 2967 &-- &-- &-- &  \multirow{2}{*}{4207}&  \multirow{2}{*}{3823}\\
 &$+$BSSE & 2761& 2557& 2161 & 3823 & 2739 & 2775 & 2710 &-- &-- &--  \\
 \hline
\multirow{2}{*}{ASW(3)}&$-$BSSE  & 2886  & 3591 & 3067 & 2449 & 4054 & 3161 & 2849 & 2260 & 3709 & -- &  \multirow{2}{*}{4054} &  \multirow{2}{*}{3660}\\
& $+$BSSE& 2684  & 3165 & 2859 & 2249 & 3660 & 2936 & 2653  & 2047 & 3446  & --  \\
\hline
\multirow{2}{*}{ASW(4)}&$-$BSSE  & 2764  & 3737 & 2640  &3784 & 3091 &3072  & 2096 & 3619 & 3137 & 2807 &  \multirow{2}{*}{3784} &  \multirow{2}{*}{3485}\\
 & $+$BSSE& 2570 & 3305 & 2422 & 3485 & 2889 & 2601  & 1904 & 3269 & 2825 & 2607  \\
 \hline
\multirow{2}{*}{ASW(5)}&$-$BSSE  & 3337 & 5364 & 2804 & 2512  & 3153 & 3229 & 3287 & 5362 &-- &-- &  \multirow{2}{*}{5364}&  \multirow{2}{*}{4937}\\
 &$+$BSSE & 3137 & 4937  & 2591 & 2330 & 2944  & 3031  & 3022 & 4935  &-- &--  \\
 \hline
\multirow{2}{*}{ASW(6)}&$-$BSSE  & 2408 &  2376  &3377 &2263 & 4834 &3874 &2927 & 3397 &1998 &-- & \multirow{2}{*}{4834} &  \multirow{2}{*}{4415}\\
  &$+$BSSE& 2181 & 2180 & 3180& 2080& 4415& 3411 & 2723 & 3163 & 1846 &--  \\
  \hline
\multirow{2}{*}{ASW(7)}&$-$BSSE  & 3127 & 2972 & 3162 &2422 & 1765 &3035 &2664 &3142 &-- &-- & \multirow{2}{*}{3162} &  \multirow{2}{*}{2953} \\
 &$+$BSSE& 2746 & 2633  & 2953 & 2224 & 1624 & 2831 & 2460 & 2759 &-- &-- \\
 \hline
\multirow{2}{*}{ASW(8)}&$-$BSSE & 3051 & 2498 & 3291 &2831 & 2327& 2519 & 2818 & 2598 & 3029 &-- & \multirow{2}{*}{3291} &  \multirow{2}{*}{2972}\\
&$+$BSSE& 2745 & 2303& 2972 & 2630 & 2142 & 2329 & 2581 & 2406 & 2793 &--  \\
\hline
\multirow{2}{*}{ASW(9)}&$-$BSSE  &2913 & 3260 & 3736 & 2926 & 4157 & 2300 & 3601 & 3248 &-- &-- &  \multirow{2}{*}{4157}&  \multirow{2}{*}{3865}\\
 &$+$BSSE& 2721 & 3072 & 3441 & 2717 & 3865 & 2003 & 3396 & 3023 &-- &-- \\
\hline
\hline
& & \multicolumn{10}{c}{ Total 75 binding sites across 9 clusters} &{ Average $-$BSSE $= 4023$} & { Average $+$BSSE $= 3696$} \\
\hline
    \end{tabular}}
  \end{table}
  {\color{black}
\begin{itemize}
    \item   Scaling factor $-$BSSE $=4023/3354= 1.200$ 
 \item  Scaling factor $+$BSSE $=3696/3154=1.172$
\end{itemize}
}
  \end{landscape}

\begin{landscape}

  \begin{table}[H]
   \centering
   \caption{{\color{black}Calculated BE values for the CO and CS considering ASW(1) (minimum) $\rm{[H_2O]_{20}}$ cluster. All values reported are in the Kelvin (K) unit.}}
   \label{tab:CO_CS}
    \resizebox{\linewidth}{!}{\begin{tabular}{c c c c c c c c c c c c c}
    \hline
    \hline
{ Species} &{ BSSE}& \multicolumn{7}{c}{ $d$-H binding sites } & \multicolumn{2}{c}{ Highest BE value} & \multicolumn{2}{c}{Scaled BE value$^{(*)}$} \\
\cline{3-9} 
& { $(-/+)$}& { $d$H1 } &{ $d$H2 } & { $d$H3 } & { $d$H4 } & { $d$H5 } & { $d$H6 } & { $d$H7 } & { $-$BSSE } &  { $+$BSSE }&  { $1.188$ for $-$BSSE } & { $1.177$ for $+$BSSE} \\
\hline
\multirow{2}{*}{\ce{CO}} &$-$BSSE & 974 & 438 & 876 & 592   & 695 & 586 & 1086 &\multirow{2}{*}{1086} &\multirow{2}{*}{811} &\multirow{2}{*}{1290} & \multirow{2}{*}{955} \\
 & $+$BSSE & 735 & 272 & 669 & 439 & 577 & 464 &  811    \\
 \hline
\multirow{2}{*}{\ce{CS}} &$-$BSSE & 2286  & 1980 &  1809 & 1827   & 2047 & 2018 & 2051 &\multirow{2}{*}{2286} &\multirow{2}{*}{2072} &\multirow{2}{*}{2716} & \multirow{2}{*}{2439} \\
 & $+$BSSE & 2072 & 1790 & 1422 & 1559  & 1817 & 1845 &  1821    \\
 \hline
    \end{tabular}}
$^{(*)}$ The scaling parameters are taken from \cite{sil2024}
  \end{table}

  \begin{table}[H]
   \centering
   \caption{{\color{black}Calculated BE values for the oxygen and sulfur-bearing analogues considering ASW(1) (minimum) $\rm{[H_2O]_{20}}$ cluster. All values reported are in the Kelvin (K) unit.}}
   \label{tab:only_dH}
    \resizebox{\linewidth}{!}{\begin{tabular}{c c c c c c c c c c c c c}
    \hline
    \hline
{ Species} &{ BSSE}& \multicolumn{7}{c}{ $d$-H binding sites } & \multicolumn{2}{c}{ Highest BE value} & \multicolumn{2}{c}{Scaled BE value} \\
\cline{3-9} 
& { $(-/+)$}& { $d$H1 } &{ $d$H2 } & { $d$H3 } & { $d$H4 } & { $d$H5 } & { $d$H6 } & { $d$H7 } & { $-$BSSE } &  { $+$BSSE }&  { $1.200$ for $-$BSSE } & { $1.172$ for $+$BSSE} \\
\hline
\multirow{2}{*}{\ce{HCO}} &$-$BSSE & 2108   & 2194 & 1713 &  1977 & 2501 & 1504  & 1749 &\multirow{2}{*}{2501} &\multirow{2}{*}{2308} &\multirow{2}{*}{3001} & \multirow{2}{*}{2705} \\
 & $+$BSSE & 1914 & 2027 & 1441 & 1768 & 2308 & 1332  &  1577    \\
 \hline
\multirow{2}{*}{\ce{CH3O}} &$-$BSSE & 3105  & 3076  & 3355 & 2738 & 3332 & 2851 & 2650 &\multirow{2}{*}{3355} &\multirow{2}{*}{3085} &\multirow{2}{*}{4026} & \multirow{2}{*}{3616} \\
 & $+$BSSE & 2853  &  2837 & 2981 & 2416 & 3085 & 2620 &  2413    \\
 \hline
\multirow{2}{*}{\ce{HCS}} &$-$BSSE & 1973  & 1972 & 1821 & 1388  & 1626  & 1868 & 2050 &\multirow{2}{*}{2050} &\multirow{2}{*}{1770} &\multirow{2}{*}{2460} & \multirow{2}{*}{2074} \\
 & $+$BSSE & 1747  & 1770  & 1409  & 1157 & 1233 & 1478 &   1700   \\
 \hline
\multirow{2}{*}{\ce{H2CS}}&$-$BSSE & 2799  & 2625 & 2655  & 2360  & 3024 & 2147 & 2269 &\multirow{2}{*}{3024} &\multirow{2}{*}{2822} &\multirow{2}{*}{3629} & \multirow{2}{*}{3307} \\
 & $+$BSSE & 2591  & 2450 & 2253 & 2125 & 2822 & 1963 & 2065     \\
 \hline
\multirow{2}{*}{\ce{CH3S}} &$-$BSSE &2636 &2339 & 3061& 2404&2880 & 2038&2575 &\multirow{2}{*}{3061} &\multirow{2}{*}{2610} &\multirow{2}{*}{3673} & \multirow{2}{*}{3059} \\
 & $+$BSSE & 2371&2102 &2610 &2057 & 2607& 1800&  2247   \\
 \hline
    \end{tabular}}
  \end{table}
 \end{landscape}

\begin{landscape}
    
\begin{table}
   \centering
   \caption{{\color{black}Calculated BE values for the oxygen and sulfur-bearing analogues considering both \textit{d}$-$H and \textit{d}$-$O on ASW(1) (minimum) $\rm{[H_2O]_{20}}$ cluster. All values reported are in the Kelvin (K) unit.}}
   \label{tab:dH_dO}
\resizebox{\linewidth}{!}{
\begin{tabular}{c c c c c c c c c c c c c c c c c c c}
    \hline
    \hline
    {Species} & {BSSE} & \multicolumn{14}{c}{\textit{d}-H and \textit{d}-O binding sites}  & {Highest value} & {Scaled value$^{(\dagger)}$} \\
    \cline{3-16}
    & {($-/+$)} & $d$H1 & $d$H2 & $d$H3 & $d$H4 & $d$H5 & $d$H6 & $d$H7 & $d$O1 & $d$O2 & $d$O3 & $d$O4 & $d$O5 & $d$O6 & $d$O7 & & \\
\hline
 \multirow{2}{*}{\ce{CH2OH}} &$-$BSSE & 2729  & 3800 & 3140 & 2338 & 4936 &  2428 & 2450 & 4927 & 4586  & 4510 & 3798 & 4202 & 4111 & 3935 & 4936  & 5923\\
 & $+$BSSE & 2434  & 3424 & 2644 & 2054 & 4568 & 2173 & 2183 & 4560 & 4025 & 4117 & 3489 & 3780 & 3757 & 3592 & 4568 &  5354    \\
 \hline
 \multirow{2}{*}{\ce{CH3OH}} &$-$BSSE &  3699&3510 &3853 & 3379 &3686 &3002 & 3323& 4694 & 5081 & 4201 & 3047 & 2958 & 3139  &  3755 & 5081 & 6097\\
 & $+$BSSE &3292 &3088 & 3293& 2980& 3277&2632 & 2924 & 4270 & 4418 & 3762 &2726 & 2629 & 2825  & 3388 & 4418 & 5178   \\
 \hline
 \multirow{2}{*}{\ce{C2H5OH}} &$-$BSSE & 3892 & 3501 & 5371& 3410&3638 & 3678& 3619& 4715 & 3233 & 4342 & 3020 & 2924 & 3003 & 2741 & 5371 & 6445  \\
 & $+$BSSE &3497 & 3118 & 4593 & 3040 & 3200& 3305 &  3250 & 4323 & 2875 & 3965 & 2693 &  2620 & 2695 & 2405 & 4593  & 5383  \\
 \hline
 \multirow{2}{*}{\ce{C3H7OH}} &$-$BSSE &3897 & 4303& 3849& 4247& 3850& 3529 & 4210 & 4873 & 5464 & 4528 & 3843  & 2904 & 3779 & 3773 & 5464  & 6557 \\
 & $+$BSSE &3525 &3812 &3441 &3838 & 3357& 3076& 3690 & 4349 & 4800 & 4086 & 3458 & 2609 & 3359 & 3422 & 4800  & 5626  \\
\hline

 \multirow{2}{*}{\ce{CH2SH}} &$-$BSSE & 2318& 2398& 3006&2777 &2410 &1545 & 3035 & 2868 & 2029 & 2594 & 2256 & 2387 & 2387 & 2435 & 3035  & 3642 \\
 & $+$BSSE & 1939& 2068 & 2401 & 2315& 2061& 1307& 2526 & 2508  & 1670  & 2223 & 1983 & 2040 & 2056 & 2095 &  2526 & 2961  \\
 \hline
 \multirow{2}{*}{\ce{CH3SH}} &$-$BSSE & 2783& 2812& 3691& 2380& 2516 & 2500& 2574 & 3021 & 3326  & 2895 & 1966 & 2730 &  1678 & 2569 & 3691 & 4429 \\
 & $+$BSSE & 2395 & 2458& 3116&2051 &2110 &2235 & 2277& 2543 & 2728 & 2420 & 1687 & 2243 & 1449 & 2195 & 3116 & 3652   \\
 \hline
 \multirow{2}{*}{\ce{C2H5SH}} &$-$BSSE & 3186& 2959 & 2414 &2705 & 2680 & 2743 & 3036 & 2515  & 3362  & 2791 &2561 & 3140 & 2527 & 2257 & 3362 & 4034\\
 & $+$BSSE & 2867 & 2591 &  2182 & 2361&2375 & 2409&  2719  & 2181 & 2950 & 2476 & 2219 & 2684 & 2163 & 1916 & 2950 & 3457   \\
 \hline
 \multirow{2}{*}{\ce{C3H7SH}} &$-$BSSE &3837 & 3489 & 4319& 3927& 3492 & 2769 & 3187 & 2845 & 4492 & 3054 & 3366 & 3689 & 3385 & 3598 & 4492  & 5390 \\
 & $+$BSSE & 3336 & 3093 & 3713 & 3446 & 2971 & 2487 &  2719   & 2419 & 3793  & 2624 & 2926 &  3234 & 2958  & 3146 & 3793 & 4445  \\
 \hline
    \end{tabular}}
    \vspace{1em}\\
    $^{(\dagger)}$  $1.200$ for $-$BSSE and $1.172$ for $+$BSSE
  \end{table}
 \end{landscape}

\end{document}